\documentclass[a4paper,11pt]{article}
\usepackage{amssymb}	
\usepackage{amsmath}
\usepackage{jinstpub}
\usepackage[utf8]{inputenc}
\usepackage{graphicx}
\usepackage{epstopdf}
\usepackage{todonotes}
\usepackage{lineno,hyperref}
\usepackage{upgreek}
\usepackage{units}
\usepackage{multirow}
\usepackage{array}	
\usepackage{textcomp}
\usepackage{pdfpages}
\usepackage{caption}
\usepackage{subcaption}
\usepackage{float}
\usepackage[section]{placeins}
\usepackage{natbib}
\usepackage{comment}
\usepackage{booktabs}
\usepackage{adjustbox}
\usepackage{lineno}

\nocite{*}

\title{Quality Concerns Caused by Quality Control - deformation of silicon strip detector modules in thermal cycling tests} 

\author[1,2]{Richard Salami,} 
\author[1,2]{Luise Poley,}
\author[3]{Kirsten Affolder,}
\author[3]{Tony Affolder,}
\author[4]{Lukas Bayer,}
\author[5]{Ben Crick,}
\author[6, 12]{Emily Duden,}
\author[7]{Ian George Dyckes,}
\author[3]{Vitaliy Fadeyev,}
\author[7]{Anne Fortman,}
\author[8]{Pavol Federi\v c,}
\author[4]{Laura Franconi,}
\author[3]{Matthew Gignac,}
\author[6]{Shubham Gupta,}
\author[4]{John Hallford,}
\author[9]{Cole Helling,}
\author[5]{Ewan Hill,}
\author[7]{Miao Hu,}
\author[8]{Ji\v r\' i Kroll,}
\author[10]{Priyanka Kumari,}
\author[11]{Carlos Lacasta,}
\author[9]{Madison Levagood,}
\author[11]{Hanlez Lopez,}
\author[3]{Len Morelos-Zaragoza,}
\author[12]{Meny Raviv Moshe,}
\author[6]{Aaron Petersen,}
\author[11]{Vicente Platero,}
\author[1,2]{Archa Devi Rajagopalan,}
\author[4]{Lisa Sitnikova,}
\author[11]{Carles Solaz,}
\author[11]{Urmila Soldevila,}
\author[1,2]{Peter Speers,}
\author[12]{Gerrit Van Nieuwenhuizen,}
\author[3]{Alex Zeng Wang,}

\affiliation[1]{Department of Physics, Simon Fraser University, University Dr W, Burnaby, Canada}
\affiliation[2]{TRIUMF, Wesbrook Mall, Vancouver, Canada}
\affiliation[3]{Santa Cruz Institute of Particle Physics, University of California, High Street, Santa Cruz, United States of America}
\affiliation[4]{Deutsches Elektronen-Synchrotron, Notkestra\ss{}e, Hamburg, Germany}
\affiliation[5]{Department of Physics, University of Toronto, Saint George St., Toronto, Canada}
\affiliation[6]{Martin A. Fisher School of Physics, Brandeis University, Waltham, United States of America}
\affiliation[7]{Lawrence Berkeley National Laboratory, Cyclotron Road, Berkeley, USA}
\affiliation[8]{Institute of Particle and Nuclear Physics, Charles University, Ke Karlovu, Prague, Czech Republic}
\affiliation[9]{University of British Columbia, Department of Physics, Agricultural Road, Vancouver, Canada}
\affiliation[10]{Department of Physics and Astronomy, York University, Keele Sreet, Toronto, Canada}
\affiliation[11]{Instituto de F\'{\i}sica Corpuscular, CSIC-Universidad de Valencia, c/ Catedr\'{a}tico Jos\'{e} Beltr\'{a}n, Paterna, Spain}
\affiliation[12]{Brookhaven National Laboratory, Rochester Street, Upton, United States of America}

\emailAdd{ros3@sfu.ca}

\abstract{The ATLAS experiment at the Large Hadron Collider (LHC) is currently preparing to replace its present Inner Detector (ID) with the upgraded, all-silicon Inner Tracker (ITk) for its High-Luminosity upgrade (HL-LHC). The ITk will consist of a central pixel tracker and the outer strip tracker, consisting of about 19,000 strip detector modules. Each strip module is assembled from up to two sensors, and up to five flexes (depending on its geometry) in a series of gluing, wirebonding and quality control steps. During detector operation, modules will be cooled down to temperatures of about \unit[-35]{\textcelsius} (corresponding to the temperature of the support structures on which they will be mounted) after being initially assembled and stored at room temperature. In order to ensure compatibility with the detector's operating temperature range, modules are subjected to thermal cycling as part of their quality control process. Ten cycles between \unit[-35]{\textcelsius} and \unit[+40]{\textcelsius} are performed for each module, with full electrical characterisation tests at each high and low temperature point. As part of an investigation into the stress experienced by modules during cooling, it was observed that modules generally showed a change in module shape before and after thermal cycling. This paper presents a summary of the discovery and understanding of the observed changes, connecting them with excess module stress, as well as the resulting modifications to the module thermal cycling procedure. }

\begin{document}
\maketitle

\section{Introduction} 

\begin{figure} [htbp]
  \centering
  \includegraphics[width=100mm]{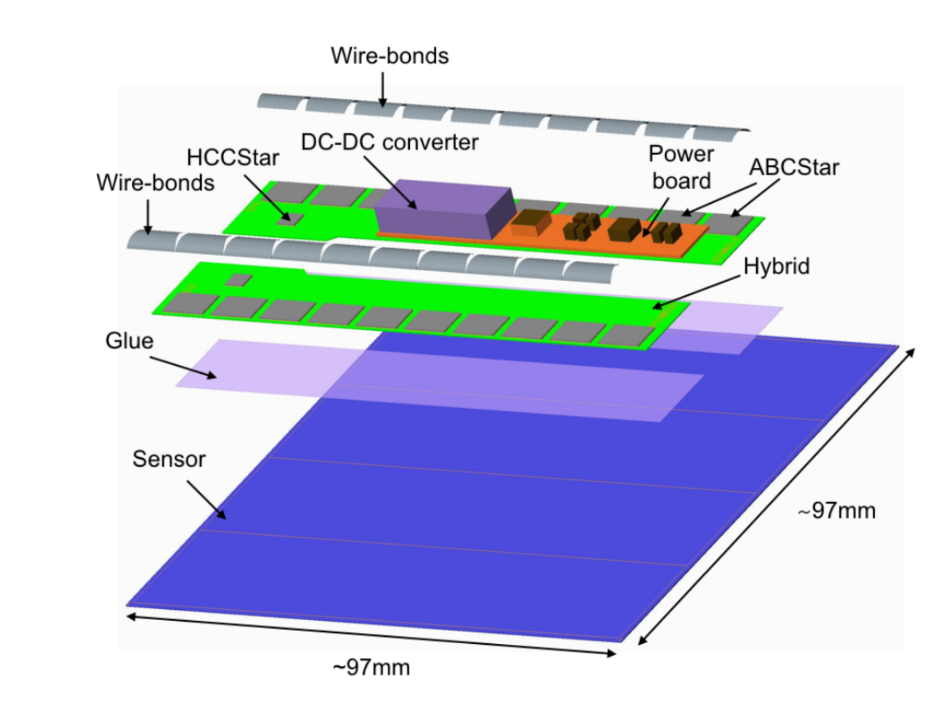}
  \caption{Exploded view of barrel module components from \cite{Tishelman-Charny_2024}.}
  \label{module-diagram}
\end{figure}

The HL-LHC upgrade introduces increased luminosity, resulting in increased radiation damage to the detector, such that the current ATLAS Inner Detector cannot be operated in the post-upgrade conditions. Thus, the Inner Detector will be replaced with the all-silicon ATLAS Inner Tracker (ITk), containing an inner pixel detector region surrounded by an outer strip detector region. The strip detector is composed of an array of modules in a central barrel region, along with two endcap regions on either side. Silicon strip detector modules, shown in Figure~\ref{module-diagram}, are constructed by gluing hybrid and powerboard flexes to the surface of silicon sensors. Strip modules are assembled in a series of assembly steps that incorporate a quality control (QC) procedure, after which the completed modules are loaded onto support structures. Among other tests, QC includes visual inspections, electrical tests, shape measurements and thermal cycling. 

Thermal cycling is the final step in QC, in which modules must operate when cycling between temperature extremes of \unit[-35]{\textcelsius} and \unit[+40]{\textcelsius} \cite{ITk-TDR}. The cycling temperatures were chosen to cover the full potential operational range in the detector.

During the strip module pre-production, it was found that modules that had passed QC had gone on to fail IV tests while mounted on support structures and cooled down. An investigation into module stress followed, with the aim of developing a better understanding of the source of the on-structure failures. The investigation into sources of stress led to the addition of module shape measurements after thermal cycling, in addition to the standard shape measurements done prior to cycling. Module shapes were found to be significantly different before and after thermal cycling. This study presents the change in module shape associated with the thermal cycling protocol and the source of module stresses prior to loading onto support structures.

\section{Modules and QC}

Modules are constructed from components in a series of manual precision assembly steps, with module component materials and thicknesses indicated in Table~\ref{component_table}. Modules in the barrel region, shown in Figure \ref{module-diagram}, are square long-strip (LS) or short-strip (SS). Modules in the endcap are approximately trapezoidal, with curved edges. Hybrids are the flexible circuit boards that hold the readout chips (ASICs) responsible for registering hit signals, as well as housing thermistors for temperature monitoring. The powerboard provides the DC-DC power for the module.

To assemble a module, the ASICs are glued and wirebonded to the hybrids, then the hybrids and powerboard are glued to the sensor, following which the hybrids and sensor are wirebonded together. 

\begin{table} [h!]
    \centering
    \resizebox{0.8\textwidth}{!}{%
    \begin{tabular}{|c|c|l|} \hline 
         \textbf{Component}& \textbf{Materia}l& \textbf{Thickness}\\ \hline \hline
         Sensor& Silicon& \unit[300-320]{\textmu m} \\ \hline 
         Hybrids (Endcap)& Copper, Polyimide& \unit[240-285]{\textmu m}\\ \hline 
 Hybrids (Barrel)& Copper, Polyimide& \unit[378-462]{\textmu m} \\ \hline 
        Powerboard (Endcap)& Copper, Polyimide& \unit[218-322]{\textmu m} \\ \hline
 Powerboard (Barrel)& Copper, Polyimide& \unit[385]{\textmu m}\\ \hline
    \end{tabular}
    }
    \caption{Module component dimensions for endcap and barrel variants \cite{metrology_proc}.}
    \label{component_table}
\end{table}

In order to ensure compatibility with the detector requirements, QC processes are performed after each module assembly step \cite{Tishelman-Charny_2024}, as shown in Table~\ref{assembly}.

\begin{table} [htbp]
\centering
\small
\begin{adjustbox}{width=\textwidth,center=\textwidth}
\begin{tabular}{|p{1.6cm}|p{2.8cm}|p{2.8cm} | p{3cm}|} \hline
\textbf{Assembly Steps} & HV tab attached to sensor & Glued & Wirebonded \\ \hline
\textbf{QC Steps} & \begin{tabular}{@{\labelitemi\hspace{\dimexpr\labelsep+0.5\tabcolsep}}l@{}}IV test\\Visual inspection\\ \end{tabular} & \begin{tabular}{@{\labelitemi\hspace{\dimexpr\labelsep+0.5\tabcolsep}}l@{}}Visual inspection\\Weighing \\Metrology \\Module shape\\IV test\end{tabular} & \begin{tabular}{@{\labelitemi\hspace{\dimexpr\labelsep+0.5\tabcolsep}}l@{}}Module shape\\IV test\\Visual inspection\\Full electrical test\\Thermal cycling\end{tabular} \\ \hline 
\end{tabular}
\end{adjustbox}
\caption{The sequence of assembly steps and the respective QC steps.}
\label{assembly}
\end{table}

\FloatBarrier{}

\begin{figure} [h!]
    \centering
    \includegraphics[width=140mm]{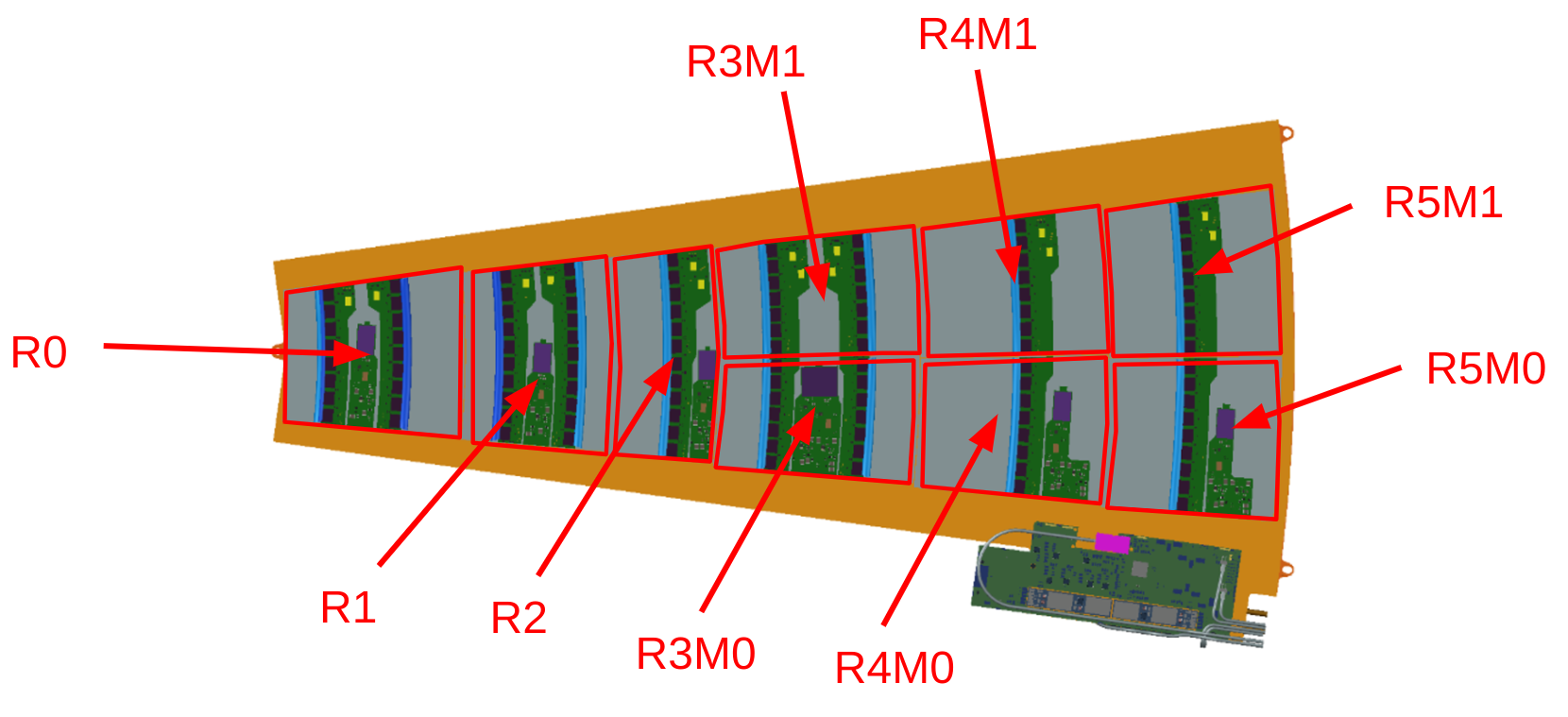}
    \caption{Schematic of a petal loaded with completed endcap modules.}
    \label{petal}
\end{figure}

Modules that pass all QC requirements are glued and wirebonded to staves or petals, shown in Figure \ref{petal}. The analyses presented in this study rely on measurements of the module shape (see section 2.1) and module leakage current tests (see section 2.2) in QC.

Hybrids and powerboards are attached to sensors using a two-component epoxy glue. The glue selected for module assembly, among other criteria, requires a short curing time, with minimal volume change to allow for precise positioning of components, as well as neutral interaction with the components to ensure good electrical performance. It must be stable under irradiation, and the glass transition temperature ($T_g$) must be significantly higher than the expected operating range of \unit[-35]{\textcelsius} to \unit[+20]{\textcelsius}, as ASICs are known to reach up to \unit[+30]{\textcelsius} higher than the module ceiling temperature of \unit[+20]{\textcelsius}.

\newpage

\subsection{Module Shape Measurement}

\begin{figure}[h!]
    \centering
    \begin{subfigure}[t]{0.48\textwidth}
        \centering
        \includegraphics[height=5cm, width=7cm]{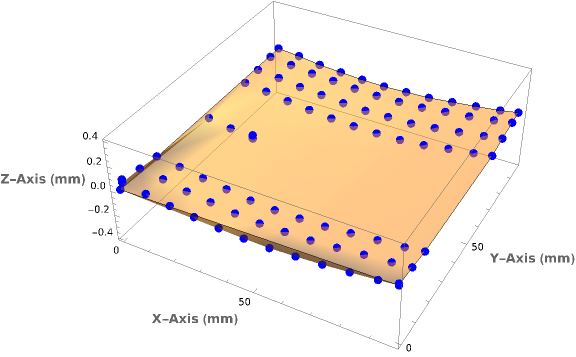}
        \caption{}
    \end{subfigure}
    \hfill
    \begin{subfigure}[t]{0.48\textwidth}
        \centering
        \includegraphics[height=5cm, width=7cm]{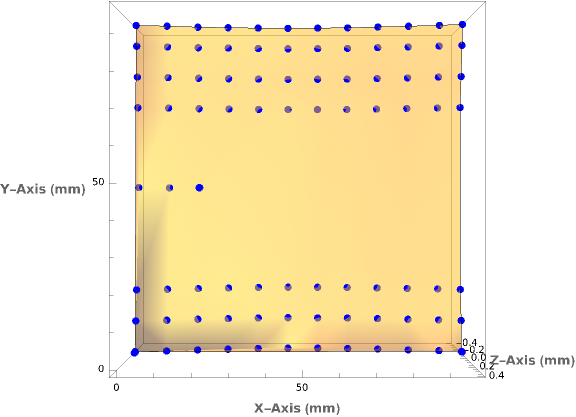}
        \caption{}
    \end{subfigure}
    \caption{Measurement grid on a barrel module. The blue points represent measurements taken in the module shape test, with the full module shape interpolated in orange. The powerboard and hybrid flexes are not measured, as they are not considered in the module shape.}
    \label{met-grid}
\end{figure}

Module shapes are measured with the module placed on a flat surface to obtain its natural shape. Out-of-plane height measurements are taken on a grid of points across the sensor surface that are at most \unit[1]{cm} apart, see Figure~\ref{met-grid}. This is the general procedure for the module shape tests.

\FloatBarrier{}

The module shape test in QC requires the module shape coefficient to measure between \unit[-50]{\textmu m} and \unit[+150]{\textmu m} \cite{ABC_prototyping} to ensure that modules will be compatible with the procedure for loading onto support structures.

\begin{figure} [h!]
     \centering
     \begin{subfigure}[b]{0.32\textwidth}
         \centering
         \includegraphics[height=3cm, width=\textwidth] {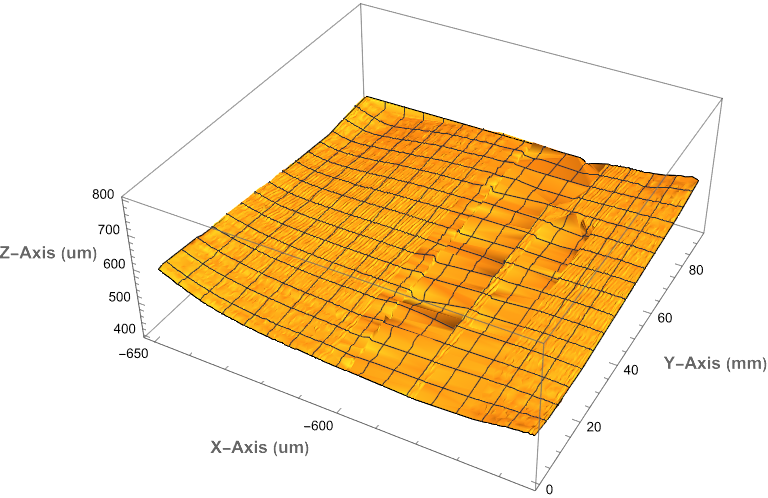}
         \caption{Orthographic view.}
         \label{tilt-ortho}
     \end{subfigure}
     \hfill
     \begin{subfigure}[b]{0.32\textwidth}
         \centering
         \includegraphics[height=3cm, width=\textwidth]{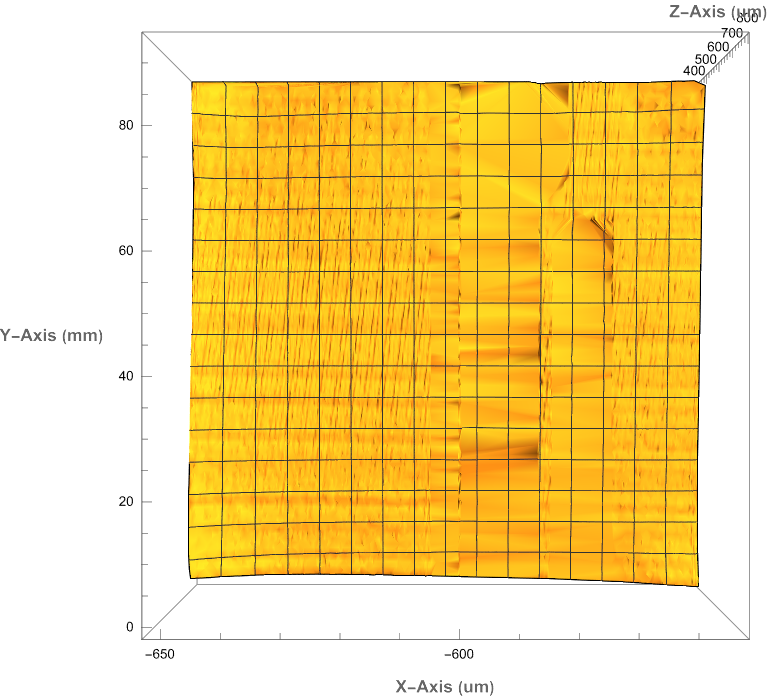}
         \caption{Top view.}
         \label{tilt-top}
     \end{subfigure}
     \hfill
     \begin{subfigure}[b]{0.32\textwidth}
         \centering
         \includegraphics[height=3cm, width=\textwidth]{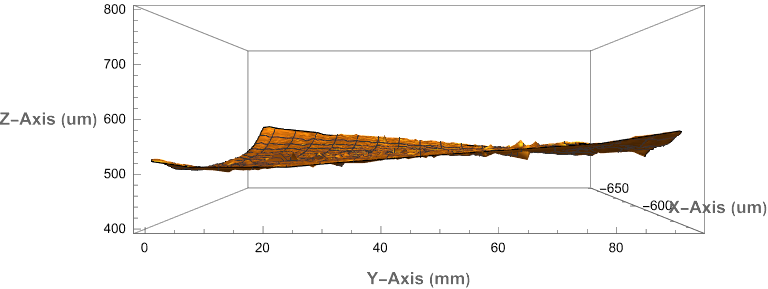}
         \caption{Side view.}
         \label{tilt-side}
     \end{subfigure}
        \caption{Tilt corrected surface map of a barrel module. The tilt corrected data yields a shape coefficient of \unit[+51]{\textmu m}. The tilt correction generally has a negligible effect on the calculated shape coefficient, suggesting that the placement of the modules during measurement does not distort the calculated shape coefficients.}
        \label{tilt-corrected}
\end{figure}

The shape measurements form a surface map of the measured module. To calculate the shape coefficient, only points on the sensor are considered (i.e. points measured on flexes or readout chips are not considered). Using all the sensor surface points, a tilt corrected surface map is obtained by using a plane fit, an example of which is shown in Figure~\ref{tilt-corrected}. The maximum height difference between points on the sensor is used as the module shape coefficient. The module shown in Figure~\ref{tilt-corrected} is a typical module that falls within the specifications for module shape coefficient. The shape of the module is consistent with the deformation introduced by the assembly process, which is acceptable for loading a module onto a support structure.

\begin{figure} [htbp]
      \centering
	   \begin{subfigure}{0.45\textwidth}
		\includegraphics[height=3cm, width=\textwidth]{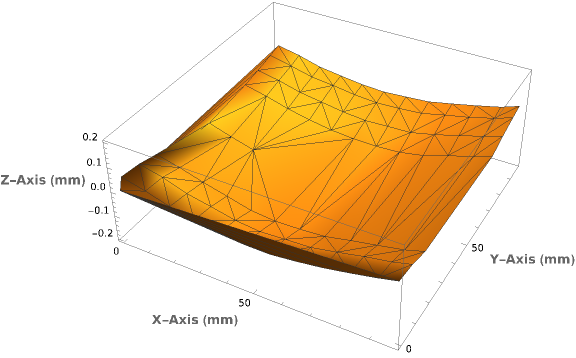}
		\caption{Positive shape coefficient, orthographic view.}
		\label{positive-shape-ortho}
	   \end{subfigure}
	   \begin{subfigure}{0.45\textwidth}
		\includegraphics[height=3cm, width=\textwidth]{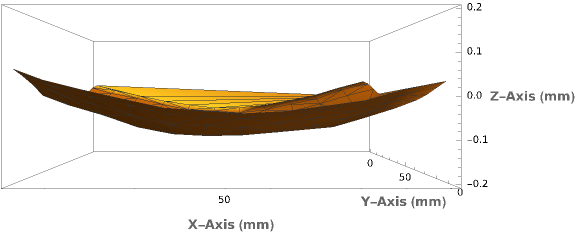}
		\caption{Positive shape coefficient, side view.}
		\label{positive-shape-side}
	    \end{subfigure}
	\vfill
	     \begin{subfigure}{0.45\textwidth}
		 \includegraphics[height=3cm, width=\textwidth]{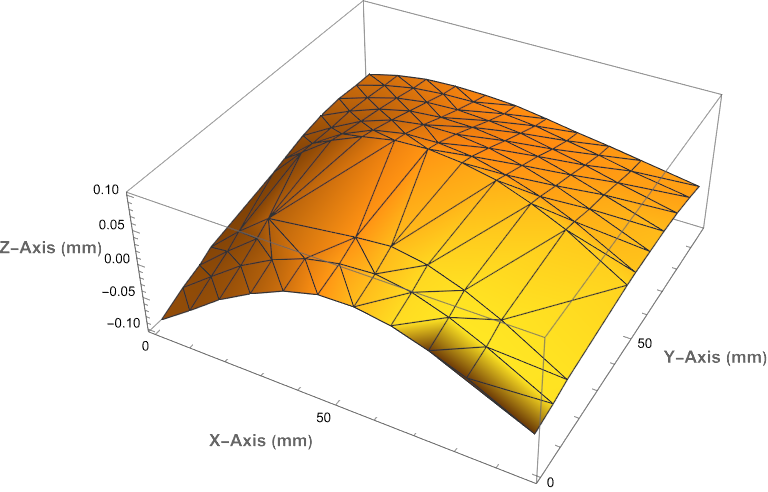}
		 \caption{Negative shape coefficient, orthographic view.}
		 \label{negative-shape-ortho}
	      \end{subfigure}
	       \begin{subfigure}{0.45\textwidth}
		  \includegraphics[height=3cm, width=\textwidth]{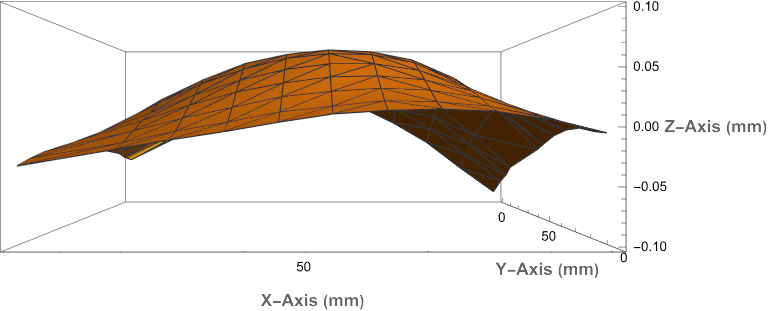}
		  \caption{Negative shape coefficient, side view.}
		  \label{negative-shape-side}
	       \end{subfigure}
	\caption{Barrel module exhibiting a positive module shape coefficient in (a) and (b) of \unit[+181]{\textmu m}. Barrel module with a negative module shape coefficient in (c) and (d) of \unit[-339]{\textmu m}. These two modules show a large bow that is outside specification for completed modules, such that they failed the shape test in QC.}
	\label{neg-pos-shape-examples}
\end{figure}

A positive module shape coefficient is defined by a global minimum for a module surface with local maxima, such that the module centre is below the edges forming a bowl shape, as shown in Figure~\ref{positive-shape-ortho} and Figure~\ref{positive-shape-side}. A negative module shape coefficient is defined by a global maximum for a module surface with local minima, such that the module centre is above the edges, forming a hill shape, as shown in Figure~\ref{negative-shape-ortho} and Figure~\ref{negative-shape-side}. 

\FloatBarrier{}

\begin{figure} [htbp]
    \centering
    \includegraphics[width=\textwidth]{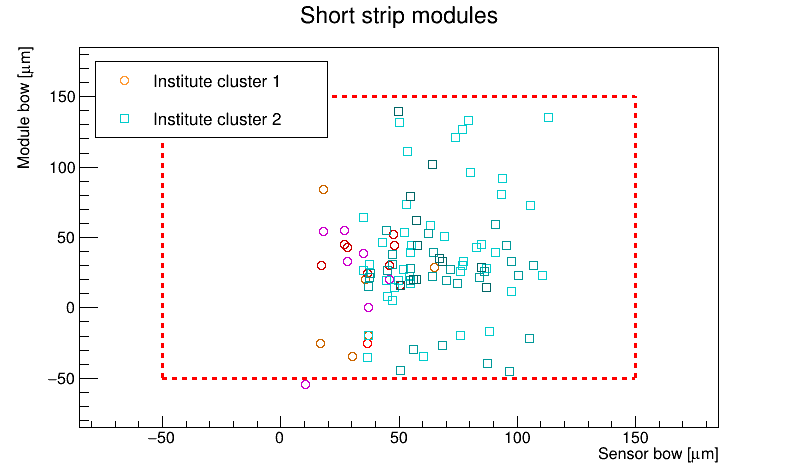}
    \caption{Sensor vs module bow (shape coefficient) for barrel SS modules.}
    \label{sensor-vs-mod}
\end{figure}

The typical module shape coefficient of a completed module is a small positive value, which is comparable to the intrinsic shape of an individual sensor in its free state. Sensor shapes are measured on bare sensors, prior to module assembly as part of a separate sensor QC regime. Intrinsic sensor deformation is known to be an effect of the sensor manufacturing process. Sensor deformation is not significantly affected by the module assembly process. Figure \ref{sensor-vs-mod} shows that the sensor and module shapes are uncorrelated, but similar in value. 

\FloatBarrier{}

\subsection{Leakage Current Measurements}

IV tests are performed between assembly steps to verify the electrical function of the modules. In an IV test, modules are subjected to voltage ramping in steps of \unit[10]{V} every 10 seconds~\cite{QC_thermal_proc}. Modules must reach the sensor bias of \unit[-500]{V} and show stability over time, as it is the nominal detector operating voltage \cite{ITk-TDR}. In order to pass the leakage current QC requirements, a module must have a measured leakage current below \unit[0.1]{$ \mathrm{\textsl{µ} A/cm^2} $}. 

\begin{figure}[h!]
    \centering
    \begin{subfigure}[t]{0.43\textwidth}
        \centering
        \includegraphics[height=5cm, width=7cm]{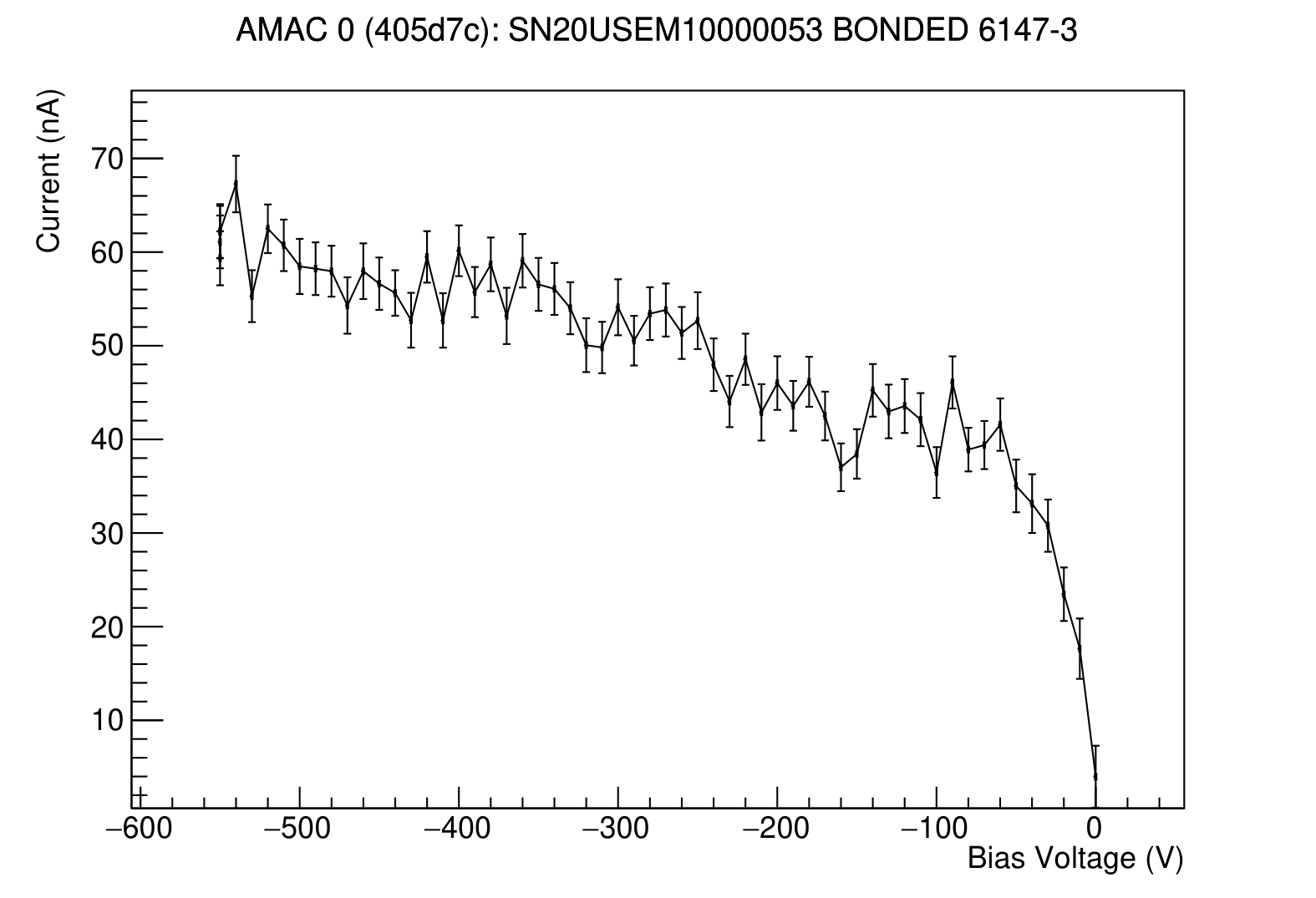}
        \caption{IV test that passes with no early breakdown.}
    \end{subfigure}
    \hfill
    \begin{subfigure}[t]{0.52\textwidth}
        \centering
        \includegraphics[height=5.1cm, width=7.5cm]{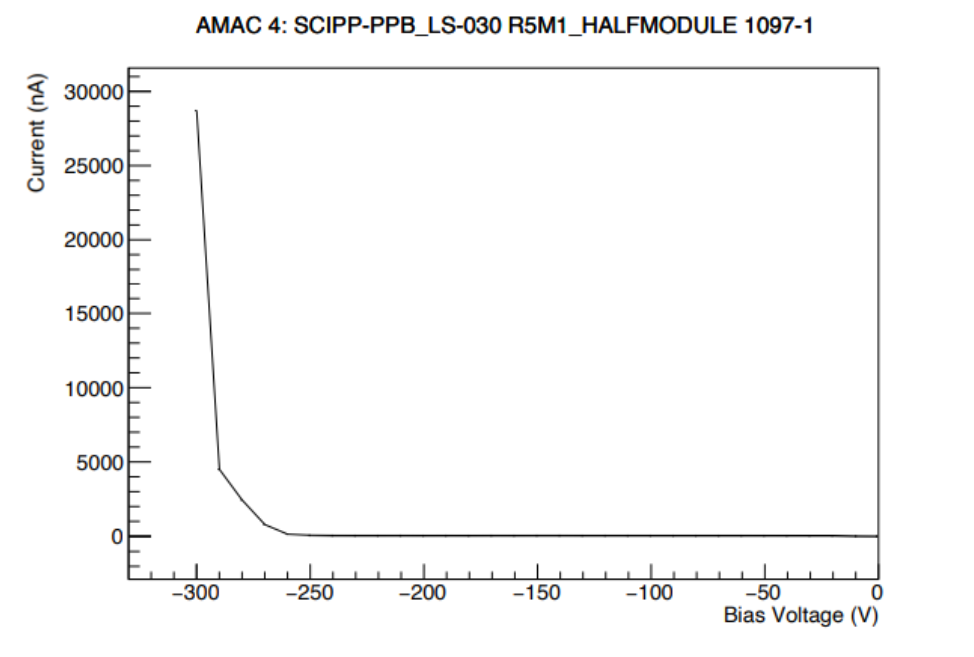}
        \caption{IV test with early breakdown at about \unit[-270]{V}.}
        \label{breakdown}
    \end{subfigure}
    \caption{IV tests for ATLAS ITk strip modules.}
    \label{IV-curve}
\end{figure}

Modules that exceed the limit for leakage current during IV tests may be experiencing early breakdown. Breakdown is a behaviour in which rapidly increasing leakage current is observed at a given sensor voltage. Early breadown is defined as breakdown occurring below the nominal operating value of \unit[-500]{V}. An example of an IV test of a module that experienced early breakdown is shown in Figure~\ref{breakdown}.

Early breakdown can indicate a variety of sensor issues such as contamination, exposure to high humidity or mechanical damage. Some occurrences of HV breakdown have been traced to sensor cracking between the powerboard and hybrids \cite{George_HSTD13}.

In addition to a QC requirement to reach bias voltages of \unit[-500]{V} for individual modules, leakage current measurements are also a powerful tool to evaluate systematic issues such as problems during the component manufacturing or module assembly process.

\FloatBarrier{}

\subsection{Thermal Cycling}

In thermal cycling, modules are evaluated through testing the electrical performance at different temperatures and ensuring the reliable module performance throughout the full operating temperature range.

\begin{figure} [h]
  \centering
  \includegraphics[width=\textwidth]{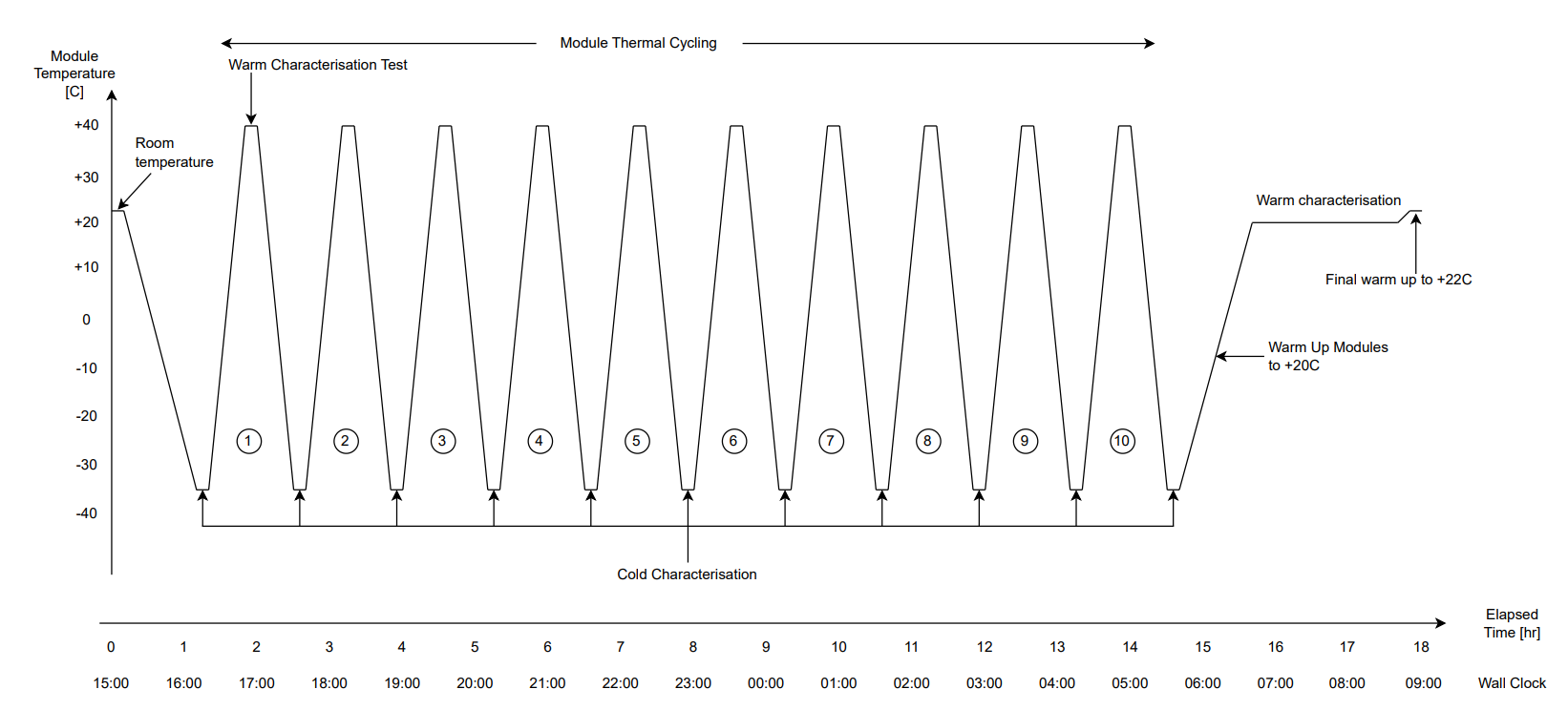}
  \caption{ATLAS ITk strips module QC thermal cycle sequence. Temperature as a function of elapsed time during a thermal cycling test, in the original cycling regime, from \cite{extreme_cycling}.}
  \label{thermal-cycle-sequence}
\end{figure}

Thermal cycling is done on multiple modules in parallel inside an insulated cold-box that is light- and air-tight (keeping relative humidity below 1\% to avoid condensation during the cycle). The cold-box holds multiple aluminum chucks, on top of which rest the test jigs. The module is screwed into an aluminum support frame that is held by the test jig. Temperature control of the module is achieved through the jig's contact with Peltier elements mounted on the chuck, in turn connected to an external chiller. The module temperature during cycling is taken to be defined by the chuck temperature. Figure~\ref{thermal-cycle-sequence} displays the temperature sequence of the thermal cycling regime.

Thermal cycling is performed in a series of steps which are identical for all modules and test sites \cite{QC_thermal_proc}:

\begin{enumerate}
  \item Initial IV test and full electrical characterization at \unit[+20]{\textcelsius}.
  \item Module power-off and initial cool down to \unit[-35]{\textcelsius}.
  \item Cold power-on test and cold IV test.
  \item Full electrical characterisation cold.
  \item HV off and temperature ramp-up.
  \item Full electrical characterization warm.
  \item Repeated ramping and tests (for 10 total cycles).
  \item Final HV stability test.
\end{enumerate}

\section{Observed Module Changes and Solutions}

After loading on staves or petals, a fraction of modules began to fail IV tests, prompting an investigation into the module properties. In module shape measurements performed on the loaded structures, it was found that modules were significantly more deformed on-structure compared to before loading, with many on-structure modules falling outside the threshold for acceptable shape coefficient. 

The observed increase in module deformation post-loading could not be caused by the loading process, as loading would only deform modules to within the acceptable threshold. Post-cycling shape measurements taken prior to loading revealed that the thermal cycling process had added to the original module deformation, represented by an increase in the average shape coefficient of about \unit[+0.2]{mm} after thermal cycling, compared to a maximum allowed shape coefficient of \unit[+150]{\textmu m}, see Figure~\ref{40C-plot}.   

The thermal cycling was identified as the cause of the increased module deformation, and the mechanism behind the change in module shape was later identified to be due to the thermo-mechanical properties of the module glue (section 3.2). 

\subsection{Module Deformation}
\begin{figure} [htbp!]
     \centering
     \begin{subfigure}[b]{0.32\textwidth}
         \centering
         \includegraphics[height=3cm, width=\textwidth]{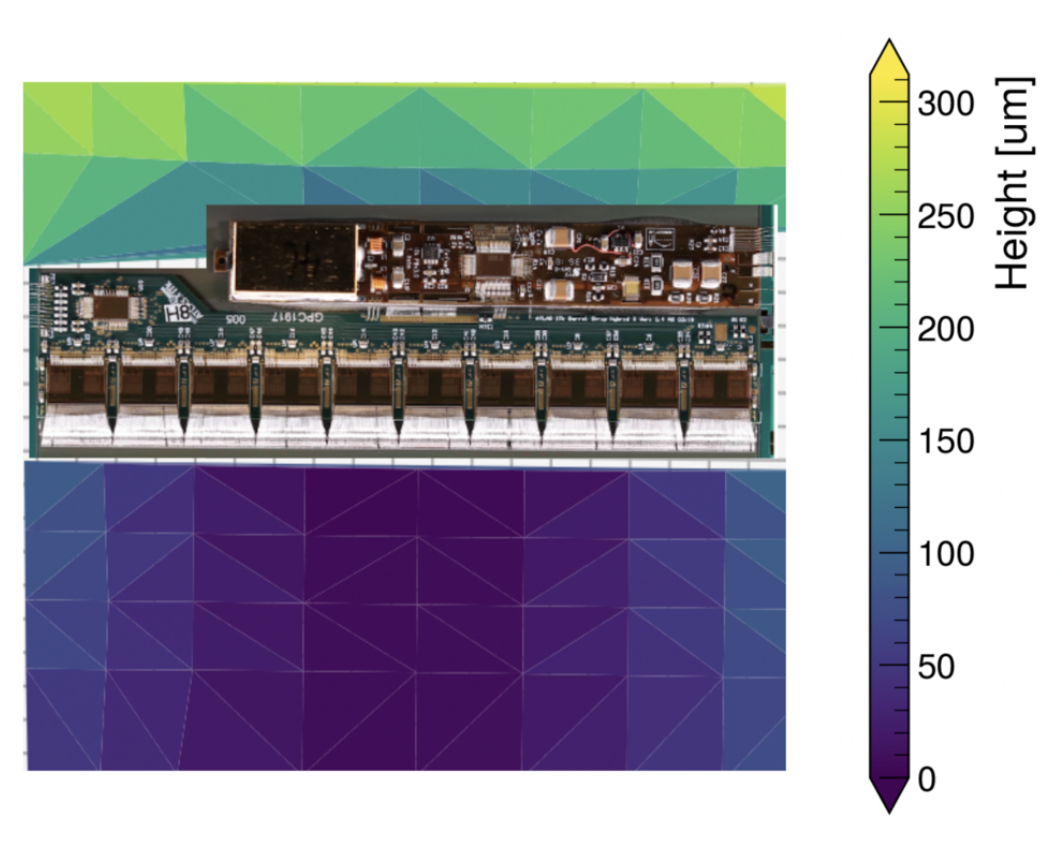}
         \caption{}
         \label{mod1-deform}
     \end{subfigure}
     \hfill
     \begin{subfigure}[b]{0.32\textwidth}
         \centering
         \includegraphics[height=3cm, width=\textwidth]{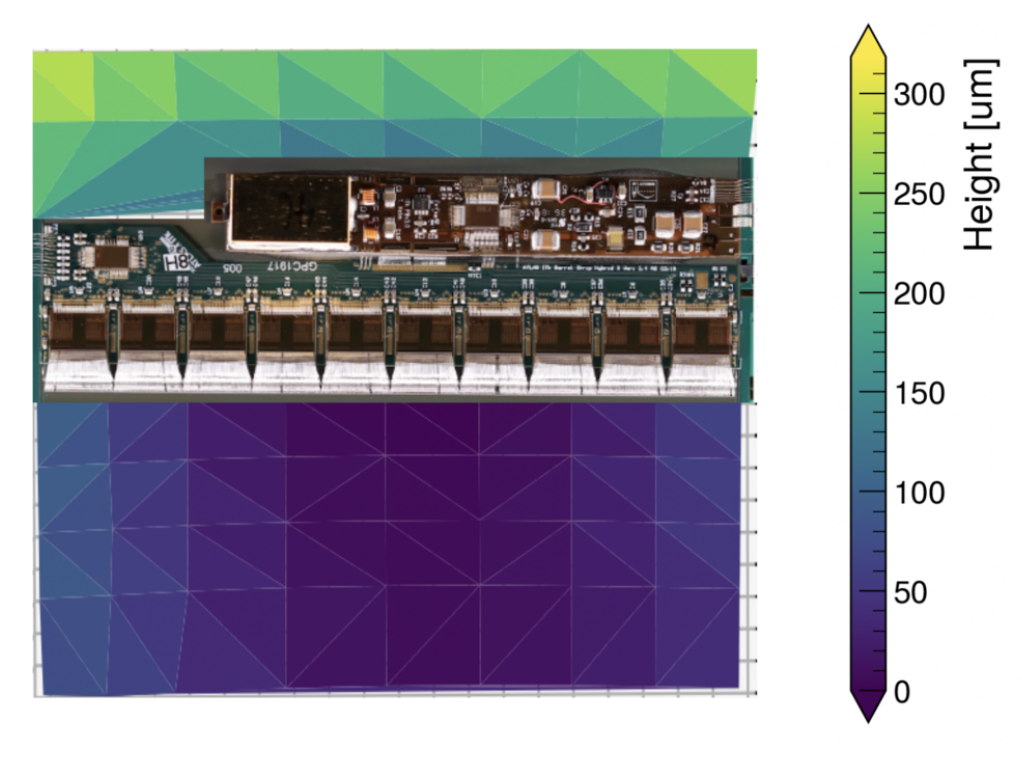}
         \caption{}
         \label{mod2-deform}
     \end{subfigure}
     \hfill
     \begin{subfigure}[b]{0.32\textwidth}
         \centering
         \includegraphics[height=3cm, width=\textwidth]{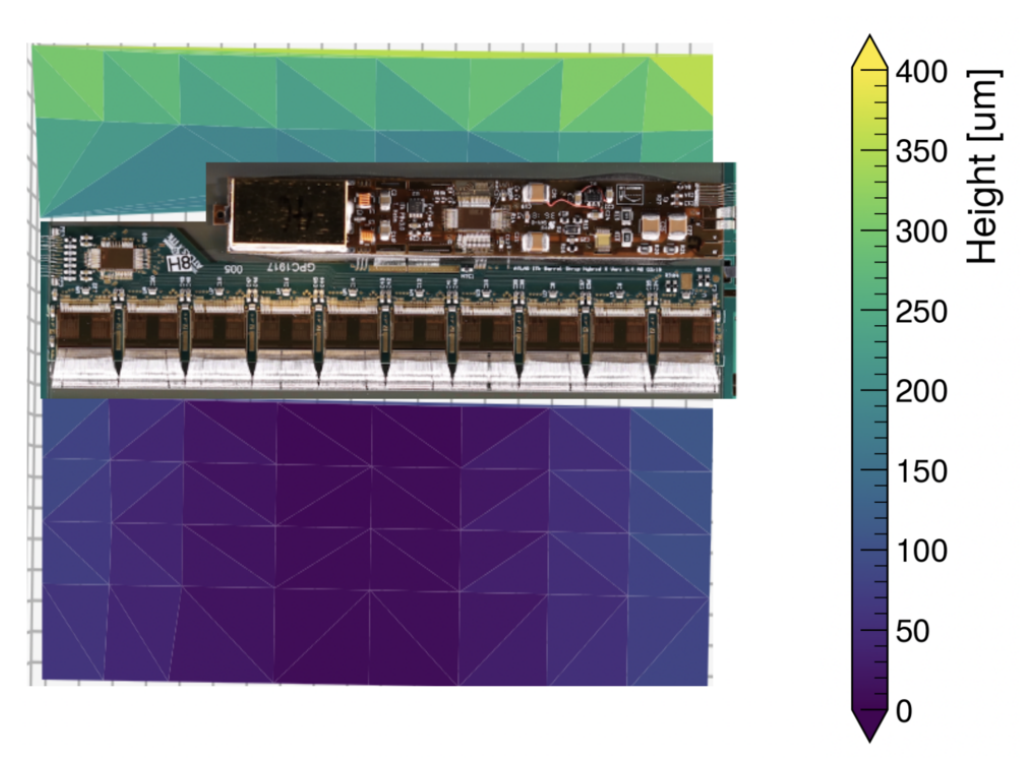}
         \caption{}
         \label{mod3-deform}
     \end{subfigure}
        \caption{Shape measurements of measured height as a function of module position for LS modules. The three modules are included in the data set used in Figure \ref{40C-plot}. The modules show an increase in shape coefficient that falls outside of the acceptable threshold. Prior to cycling, all three modules had shape coefficients within the threshold. Diagrams from \cite{extreme_cycling}.}
        \label{deformed-mod}
\end{figure}

Identifying the cause of deformation is important, as deformation introduces additional stress on modules that may lead to cracking \cite{George_HSTD13}. The module shape is known to be temperature dependent due to different coefficients of thermal expansion (CTE) between the materials of the components, causing mechanical stress on the sensor in the area between the powerboard and hybrids. Post-cycling, modules were observed to have experienced deformations, shown in Figure \ref{deformed-mod}. Instances of sensor cracking in the region between the powerboard and hybrids have been observed and are believed to be caused by stress related to the CTE mismatch \cite{Abe_CTE}.

\begin{figure} [htbp]
  \centering
  \includegraphics[width=100mm]{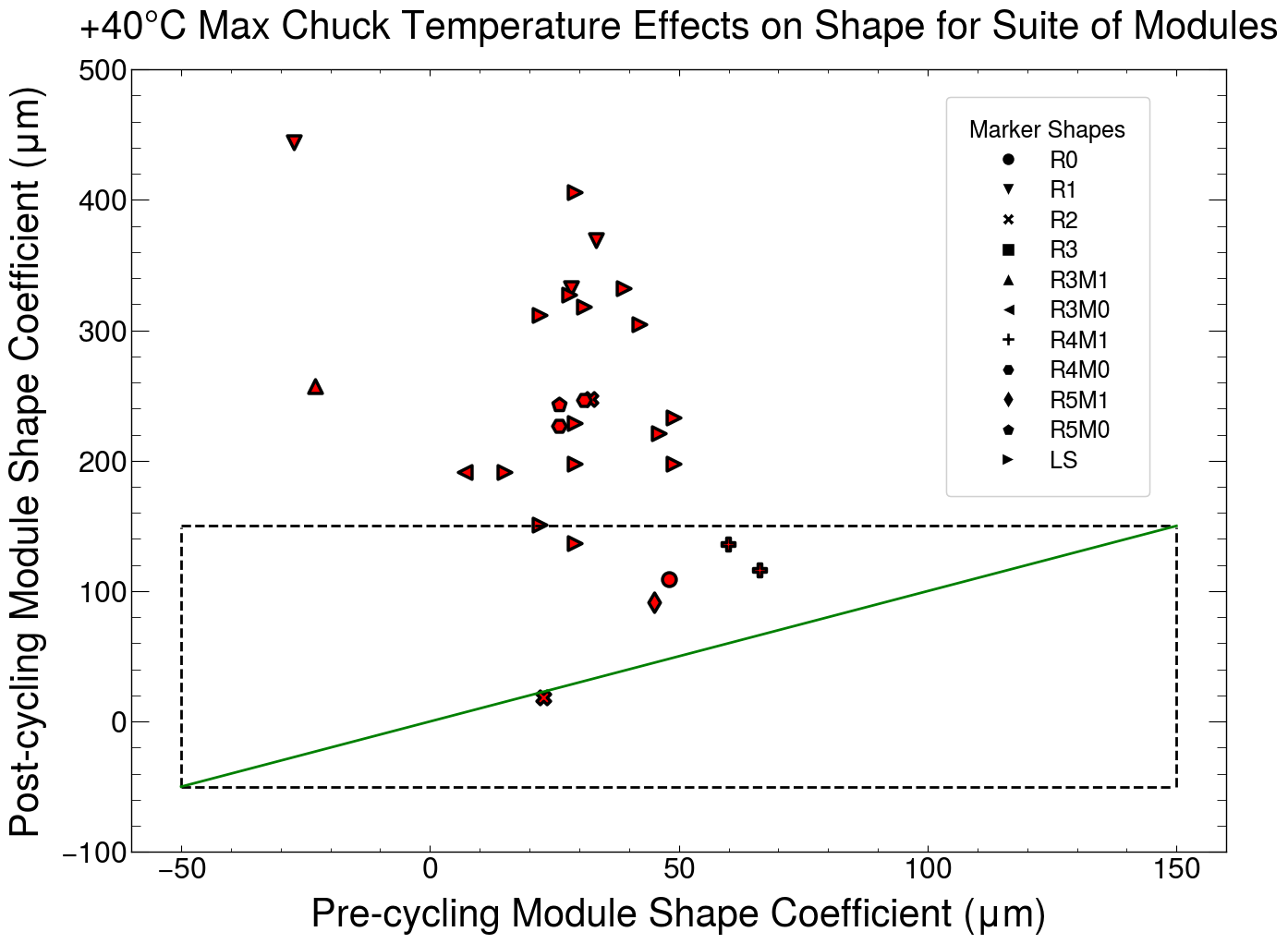}
  \caption{Post-cycling shape for modules thermally cycled to \unit[+40]{\textcelsius} ceiling. The dotted lines indicate the thresholds for the acceptable shape coefficient for modules to proceed to loading. Modules that fall upon the green diagonal line have their module shape unchanged before and after thermal cycling.}
  \label{40C-plot}
\end{figure}

The plot of post-cycling shape coefficient against pre-cycling shape coefficient is shown in Figure \ref{40C-plot}. Although these modules were within specifications before thermal cycling, 78\% of the modules in the sample exceeded the upper threshold for acceptable shape coefficient post-cycling.  

\FloatBarrier

\subsection{Glue Glass Transition} 

Previous studies observed that some temporary module deformation is always present with changes in temperature (both increasing and decreasing)\cite{capocasa_electrical_2023}, however, modules would recover their shape as they returned to room temperature.

It was thought that a higher ambient temperature during assembly correlated to higher module deformation when the module cooled, either when it returned to room temperature or was cooled in the cold-box. In order to follow up on this mechanism, thermal imaging tests were performed to check if the module temperature could become high enough to cause this effect. 

\begin{figure} [h!]
  \centering
  \includegraphics[width=110mm]{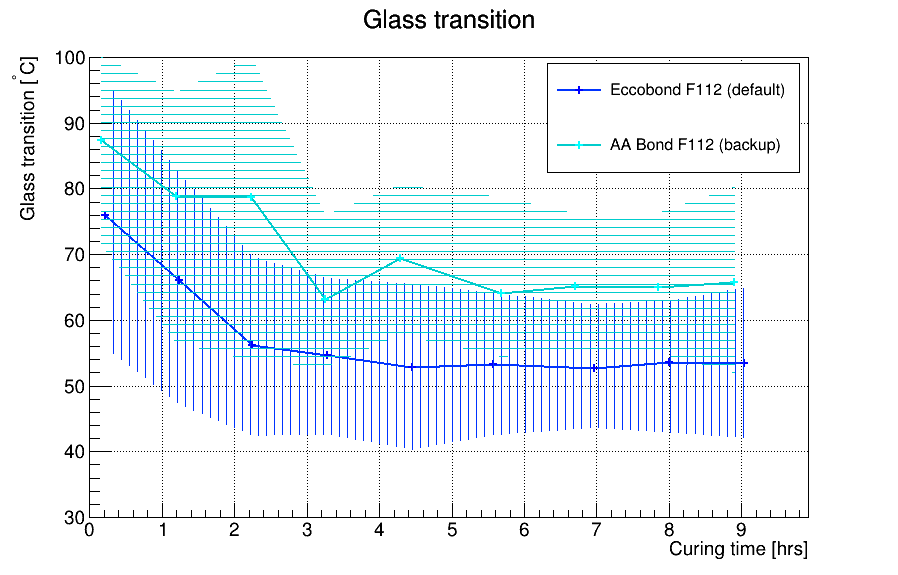}
  \caption{$ T_g $ of Eccobond F112 as cured for module assembly, measured using a Differential Scanning Calorimeter.}
  \label{glass-trans}
\end{figure}

The quoted $T_g$ of the glue, Eccobond F112, used in bonding the flexes to the sensor is around \unit[+100]{\textcelsius} \cite{eccobond_f112} for the nominal curing procedure. However, the measured $T_g$ was about \unit[+55]{\textcelsius}, falling within the temperature range of a powered module. The lower $T_g$ can likely be attributed to the non-standard curing process, as instead of applying the recommended post-cure at elevated temperature, module curing occurs at room temperature, which is known to affect the final thermo-mechanical glue characteristics. Figure~\ref{glass-trans} shows the $ T_g $ of the glue cured in this fashion, measured using standard techniques of Differential Scanning Calorimetry\cite{calorimetry}.  

\begin{figure} [htbp]
  \centering
  \includegraphics[width=110mm]{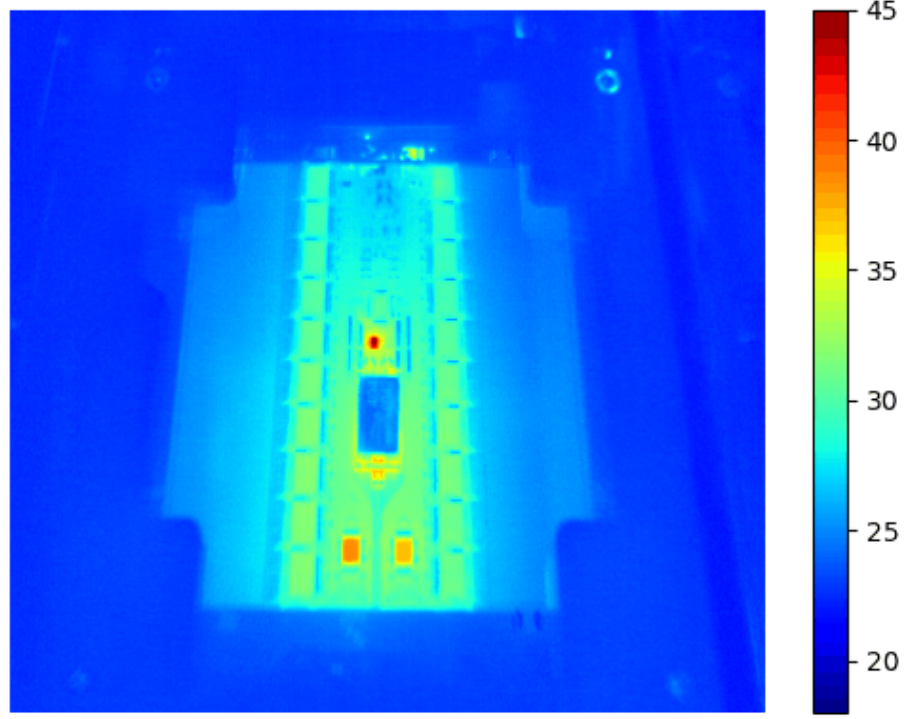}
  \caption{Thermal image of a barrel module operated at room temperature. The temperature scale is measured in \textcelsius. The highest temperature regions are around the powerboard and hybrids, as expected, the highest temperature points in these regions are the linPOL and HCCs (small chips with high power density). The overall flex temperatures are \unit[27]{\textcelsius} above the nominal chuck temperature}
  \label{thermal-image}
\end{figure}

The absolute powerboard temperature was measured at more than \unit[+45]{\textcelsius} in room temperature environment operation, as shown in Figure~\ref{thermal-image}. This suggests that in thermal cycling to \unit[+40]{\textcelsius} chuck temperature, the absolute module temperature exceeds \unit[+45]{\textcelsius} around the powerboard.

\begin{figure} [htbp]
  \centering
  \includegraphics[width=120mm]{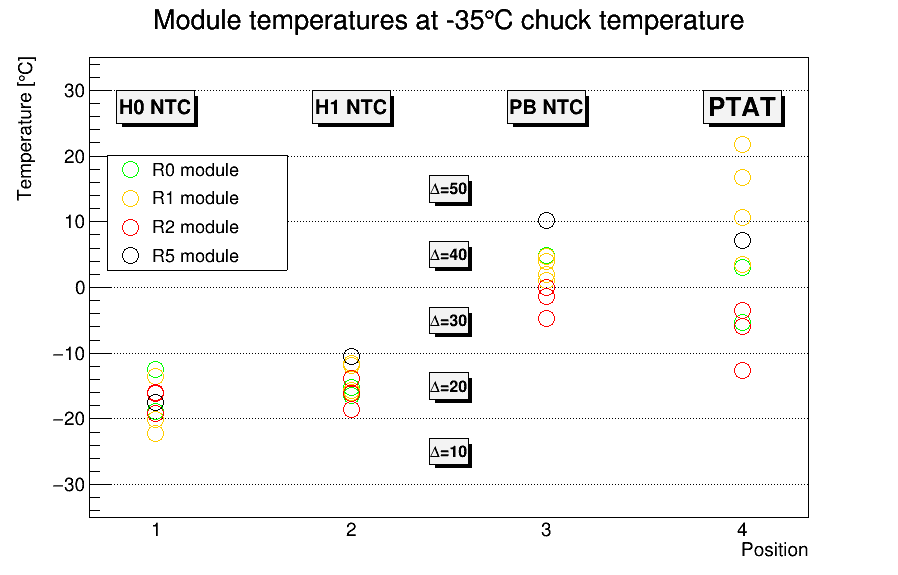}
  \caption{Module temperature measured at the cold floor (chuck temperature of \unit[-35]{\textcelsius}) of the thermal cycle for endcap modules. The temperatures are measured at the positions of the thermistors on the hybrids (H0 and H1 NTCs) and on the powerboard (PB NTC). The temperature as measured by the proportional to absolute temperature channel (PTAT) is also reported. The y-axis measures the absolute module temperature, with the temperature deltas showing the module temperature relative to the chuck.}
  \label{flex-temps}
\end{figure}

Measurements of module temperatures when in the cold stage of the thermal cycle show that the thermistor on the hybrid and powerboard flexes measures a relative temperature between \unit[50]{\textcelsius} to \unit[70]{\textcelsius} above the chuck temperature, see Figure~\ref{flex-temps}. The large temperature difference between the module and the chuck is what allows the module temperature to surpass the glue $ T_g $ once the chuck reaches the temperature ceiling.

\begin{figure} [htbp]
  \centering
  \includegraphics[width=110mm]{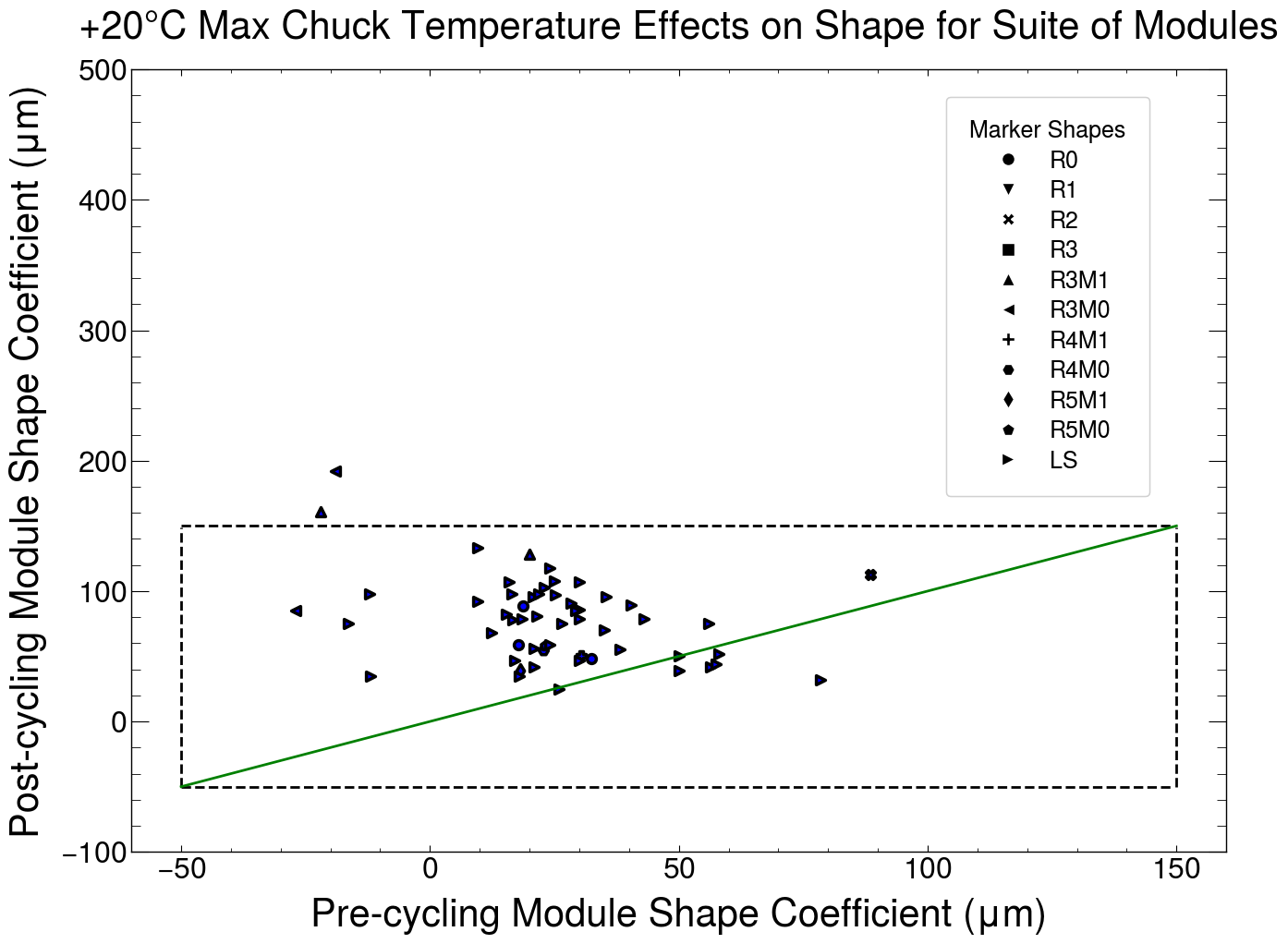}
  \caption{Post-cycling shape coefficient for modules thermally cycled to \unit[+20]{\textcelsius} ceiling. Most modules fall near the green line, indicating the shape coefficient is similar before and after thermal cycling.}
  \label{20C-plot}
\end{figure}

Following the discovery of the glue's $T_g$ being significantly lower than the nominal value, it was suspected that the thermal cycles were hot enough to be surpassing the glue's $T_g$ and inducing the post-cycling deformation. In order to verify this assumption, a series of module cycling tests was performed with a lower ceiling temperature of \unit[+20]{\textcelsius} to check the impact on the module shape. The resulting plot for the modified protocol is shown in Figure \ref{20C-plot}.

The sample of modules that were thermally cycled to \unit[+20]{\textcelsius} had minimal change in shape post-cycling, with 63\% of modules having similar shape coefficients before and after cycling. The non-standard cure procedure for the glue had lowered the glue $ T_g $ to a temperature that could be exceeded at the thermal cycle ceiling of \unit[+40]{\textcelsius}. Thus, thermal cycling to \unit[+40]{\textcelsius} had allowed the absolute module temperature to exceed the glue $ T_g $ and caused a post-curing of the modules at an elevated temperature. 


Modules cycled with the modified protocol lacked permanent deformation and had post-cycling shapes measured within specification. Based on this analysis, the updated thermal cycling regime brings modules to a maximum chuck temperature of \unit[+20]{\textcelsius} in order to prevent deforming the modules in QC. The modified ceiling temperature of \unit[+20]{\textcelsius} is representative of the full range of temperatures planned during detector operation.

\FloatBarrier{}

\subsection{Direct Measurements}

\begin{figure} [htbp]
      \centering
	   \begin{subfigure}{0.45\textwidth}
		\includegraphics[height=3cm, width=\textwidth]{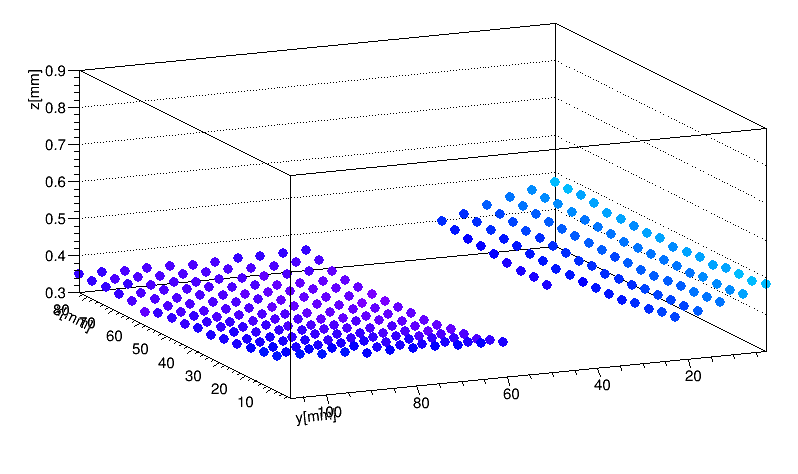}
		\caption{Module held at \unit[+30]{\textcelsius}, measuring a shape coefficient of \unit[+158]{\textmu m}}
		\label{30C-bake}
	   \end{subfigure}
	   \begin{subfigure}{0.45\textwidth}
		\includegraphics[height=3cm, width=\textwidth]{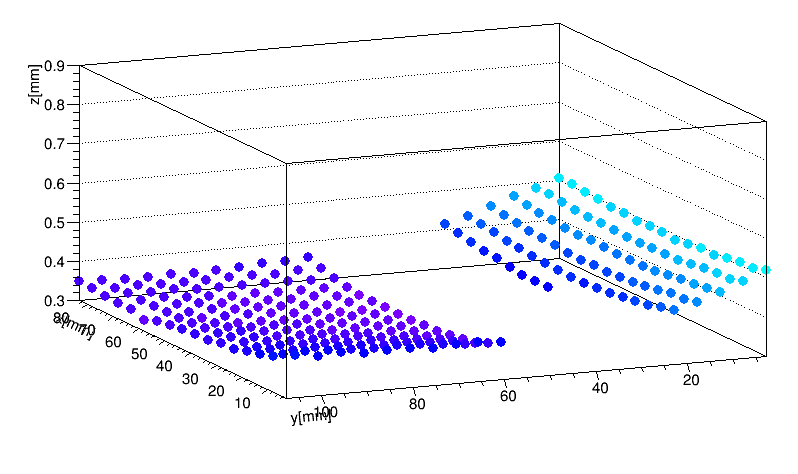}
		\caption{Module held at \unit[+40]{\textcelsius}, measuring a shape coefficient of \unit[+200]{\textmu m}}
		\label{40C-bake}
	    \end{subfigure}
	\vfill
	     \begin{subfigure}{0.45\textwidth}
		 \includegraphics[height=3cm, width=\textwidth]{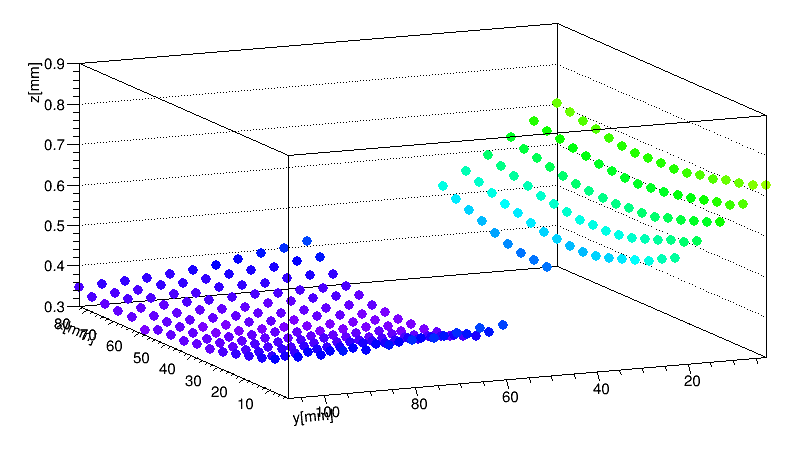}
		 \caption{Module held at \unit[+50]{\textcelsius}, measuring a shape coefficient of \unit[+420]{\textmu m}}
		 \label{50C-bake}
	      \end{subfigure}
	       \begin{subfigure}{0.45\textwidth}
		  \includegraphics[height=3cm, width=\textwidth]{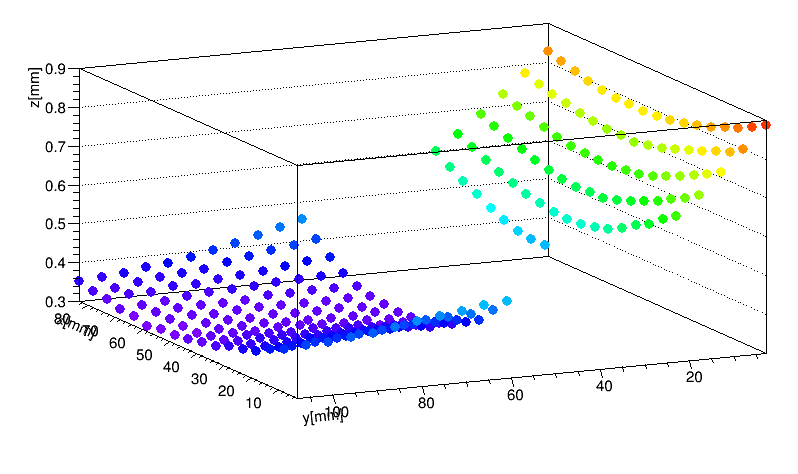}
		  \caption{Module held at \unit[+60]{\textcelsius}, measuring a shape coefficient of \unit[+580]{\textmu m}}
		  \label{60C}
	       \end{subfigure}
	\caption{Module shape measurements taken following exposure to increased temperatures.}
	\label{baketests}
\end{figure}

To confirm the effect of high temperatures on module deformation, a module was held at different elevated temperatures, allowed to cool to room temperature, and the module shape coefficient was measured following the temperature cycle. The module was held at the maximum temperatures for at least 30 minutes, with the module shape measurement done at \unit[+21]{\textcelsius}, shown in Figure \ref{baketests}. For these tests, the module was not mechanically constrained, and remained unpowered through the whole cycle, such that the module temperature would be uniform across all components.

Monitoring the module shape over a range of increased temperatures showed that temperatures above \unit[+40]{\textcelsius} led to a permanent module deformation towards higher shape coefficients. This measurement confirmed that the adhesive's glass transition temperature is lower than quoted in the material data sheet and that surpassing it can lead to module deformation.

\section{Conclusions}

As part of an investigation into early breakdown in strip modules, module shape measurements were analysed, finding that modules on staves and petals were deformed in comparison to modules pre-loading. 

The $ T_g $ of the glue with the preparation method used was much lower than the quoted value, which initiated the mechanism for the post-cycling deformation. The thermo-mechanical mechanism was found to be the heating of the module to temperatures above the glue glass transition temperature, which effectively raised the temperature at which the module was assembled, leading to larger deformation at lower temperatures, corresponding to an average module shape coefficient increase of about \unit[+205]{\textmu m}.

Investigating the source of deformation concluded that thermal cycling to a maximum chuck temperature of \unit[+40]{\textcelsius} was responsible for permanent module deformation. The problem of thermal cycling induced module deformation was resolved by reducing the thermal cycling ceiling temperature from \unit[+40]{\textcelsius} to \unit[+20]{\textcelsius}.

Reducing the cycling temperature in QC is an acceptable change, which retains the test of the full operating range, as during detector operation, the maximum operating temperature will be \unit[+20]{\textcelsius}. Following the change in thermal cycling procedure, no additional module deformation was observed.

\section*{Acknowledgements}

This work has been supported by the Canada Foundation for Innovation, the Natural Sciences and Engineering Research Council of Canada. The work at SCIPP was supported by the US Department of Energy, grant DE-SC0010107.

\nocite{*}
\bibliographystyle{jhep}
\bibliography{bibliography}

\end{document}